\documentstyle[12pt]{article}
\begin{document}
\title{
\begin{flushright}
{\small SMI-3-98 }
\end{flushright}
\vspace{2cm}
Matrix Theory in Curved Space\thanks{Talk presented at the Birthday
Conference dedicated to A.Arvilski, February, 1998}}
\author{
I.Ya. Aref'eva
  and  I.V. Volovich
\\
$~$\\
{\it Steklov Mathematical Institute}\\
{\it Gubkin St. 8, GSP-1, 117966, Moscow, Russia} \\
 arefeva, volovich@mi.ras.ru
}
\date {$~$}
\maketitle
\begin {abstract}
According to the Matrix theory proposal of Banks,
Fischler, Shenker and Susskind M-theory in the infinite momentum
frame is the large N limit of  super Yang-Mills
theory in a {\it flat} background. To address some physical issues
of classical gravity
such as gravitational collapse and cosmological expansion we consider
an extension of the BFSS proposal by defining M-theory in curved space
as the large N limit of super Yang-Mills theory in a {\it curved}
background. Motivations and possible implications of this extension
are discussed.
\end{abstract}

\newpage

It was suggested that there is a consistent quantum theory
underlying superstring theories and eleven dimensional supergravity
called M-Theory \cite{HT,Wit}.
According to the Matrix theory proposal of Banks,
Fischler, Shenker and Susskind (BFSS) \cite {BFSS} M-theory in the
infinite
momentum frame is the large N limit of the  supersymmetric Yang-Mills
theory in a {\it flat} background.
It seems that there are various M-theories
depending on the curved classical supergravity background.
The most ambitious project would be  a derivation of all possible
classical backgrounds just from the   BFSS Matrix theory, for  recent
reviews see \cite {Ban, BS, Tay}.

In this talk we discuss a less ambitious proposal.
Instead of the  derivation of the classical background from  Matrix
theory we consider Matrix theory in a given
classical background. This proposal was previously considered
in \cite{Vol}.

Such a consideration seems natural if we want to address some physical
issues of classical gravity such as gravitational collapse and
cosmological
expansion.
Scattering of branes with the large impact parameter
was successfully considered in Matrix theory in flat space
\cite{BFSS,CT,Kle,BBPT,KT,Rey}. However
it is not clear how to deal in Matrix theory  in flat space
with scattering
of 0-brane off black hole with small impact parameter which is a
characteristic process for black holes.

We discuss an extension of the BFSS proposal by defining
M-theory in curved space as the large N limit of super Yang-Mills theory
in a {\it curved} background.
Another approach to Matrix theory
in curved space is considered in \cite {DOS,BHO,Dor}.

It is well known that by using the simple string
Lagrangian
\begin {equation}
   \label {1.1}
S=\int d^2\sigma \partial_{\alpha}X^{\mu}\partial^{\alpha}
X^{\nu}\eta_{\mu\nu}
\end   {equation}
where $\alpha,\beta=0,1,~\mu,\nu=0,...,D-1$ and $\eta_{\mu\nu}$
is a flat metric in Minkowski spacetime one can  reproduce
tree scattering amplitudes in (super)gravity \cite{GSW}.
So, in string theory we can compute small corrections to the flat
Minkowski
background. However this remarkable fact
does not mean  that we are able {\it to derive} a nontrivial curved
background such as a black hole from superstring theory in flat
spacetime.
To compute amplitude for black hole creation we have to start
with an appropriate curved background, for a discussion of
this point see \cite {AVV,FVZ,Das}.
If we want to deal with string theory in a curved background with
nontrivial metric $g_{\mu\nu}$ we have to use a Lagrangian
containing this metric
\begin {equation}
\label {1.2}
S=\int d^2\sigma \partial_{\alpha}
X^{\mu}\partial^{\alpha} X^{\nu}g_{\mu\nu}
\end  {equation}
For the
theory to be self consistent, i.e. conformal invariant,  the metric
$g_{\mu\nu}$ should be Ricci flat,  $R_{\mu\nu}=0$.

The Lagrangian  (\ref{1.1}) is very simple but the
procedure of derivation of supergravity amplitude is not so simple,
it includes string  perturbation theory.

Banks, Fischler, Shenker and Susskind \cite {BFSS} have suggested
that the $U(N)$-invariant Lagrangian
\begin{eqnarray}
\label{1.3}
L=\frac{1}{2}tr[\dot{ Y}^{i}\dot{Y}^i+\frac{1}{2}[Y^i,Y^j]^2
+2\theta ^T\dot{\theta}+2\theta ^T\gamma_i[\theta ,Y^i]]
\end{eqnarray}
can be used to describe eleven dimensional
supergravity in the infinite momentum frame if one takes the large $N$
limit.
Here $Y^i$ are Hermitian $N\times N$  matrices while $\theta$
is a 16-component fermionic spinor each component of which
is an Hermitian $N\times N$ matrix and $i,j=1,...,9$.

Although the Lagrangian (\ref {1.3}) looks more complicated
than the string Lagrangian (\ref {1.1}) but there is a hope that it
has an advantage
leading to a non-perturbative formulation of quantum gravity.
This remarkable proposal does capture the essential
degrees of freedom of quantum gravity. It has passed
several nontrivial tests including the derivation
of the effective action for the long-distance and
low-energy scattering.

However this remarkable fact
does not  mean  that we are already able to  derive a nontrivial
background
such as a black hole with its nontrivial topology from the Matrix theory
Lagrangian (\ref {1.3}).  There are skilful constructions
of  branes in Matrix theory as operator matrices of special form
but they are lacking the crucial global property of black hole,
i.e. its horizon.

We treat the Matrix theory Lagrangian (\ref{1.3}) as an
analogue of the String Lagrangian  (\ref {1.1}) in flat Minkowski
spacetime.  Indeed  the Lagrangian (\ref {1.3})
can be regarded as $U(N)$ supersymmetric Yang-Mills theory
in  ten-dimensional {\it flat} Minkowski spacetime dimensionally reduced
to
$(0+1)$ space-time dimensions \cite{CH,DFS,KP}.
Therefore if we want to deal with Matrix theory in a nontrivial
curved background with metric $g_{\mu\nu}$ then by analogy with
(\ref{1.2})
we have to consider  instead of (\ref{1.3}) another Lagrangian
which is obtained from $U(N)$ supersymmetric Yang-Mills theory
in  ten-dimensional {\it curved} spacetime dimensionally reduced to
$(0+1)$ space-time dimensions.  The bosonic part of the obtained
Lagrangian reads \cite{Vol}
\begin{eqnarray}
\label{1.4}
L=-\frac{1}{2}tr[\dot{Y}_i\dot{Y}_jg^{00}g^{ij}
-\frac{1}{2}[Y_i,Y_j][Y_m,Y_n]g^{im}g^{jn}]
\end{eqnarray}
Here $g^{\mu\nu}=g^{\mu\nu}(t)$ are functions
of time $t$.
The Lagrangian (\ref{1.4}) is reduced to the bosonic part
of (\ref{1.3})
if one takes $g^{00}=-1,~g^{ij}=\delta^{ij}$.

Notice that we would obtain the dependence on time
even if we consider a static metric. This is because
one can take the dimensional reduction of the Yang-Mills
theory to a geodesic in curved space.
In this case functions
$g^{\mu\nu}$ depend from the parameter on the geodesic.

If we want to treat  (\ref{1.4}) as an extension
of the Matrix theory Lagrangian (\ref{1.3}) then
we should interpret $g^{\mu\nu}(t)$ not as a spacetime  metric
but just as functions of parameter $t$ because
spacetime is dynamically generated as a collective
mode being constructed from matrices $Y_i$.

Classical dynamical system  (\ref{1.3}) has been discussed in
\cite{AMRV}. Properties of the dynamical system (\ref{1.4})
are different. For example for the Kasner metric

\begin{eqnarray}
\label{1.5}
ds^2=-dt^2+\sum_i t^{2p_i}dx_i^2
\end{eqnarray}
for the  ansatz $Y_1=y_1\sigma_1, Y_2=y_1\sigma_2$
one has equations

\begin{eqnarray}
\label{1.5a}
\frac{d}{dt}(t^{-2p_1+1}\dot {y}_1)+t^{-2p_1-2p_2+1 }y_1y_2^2=0,~~
\frac{d}{dt}(t^{-2p_2+1}\dot {y}_2)+t^{-2p_1-2p_2 +1}y_2y_1^2=0
\end{eqnarray}

The low energy effective theory of D-branes in Minkowski spacetime
is given by the dimensional reduction of the supersymmetric
gauge theory in ten
dimensional Minkowski
spacetime \cite{Wit2}. If one has D-branes
in {\it flat } Minkowski spacetime but in  {\it curved} (non-Cartesian)
coordinates we have to start from the supersymmetric Yang-Mills theory
in
curved coordinates.  Then we get  a version of the M(atrix) theory
Lagrangian (\ref{1.4}) in the curved coordinates.  If one has D-branes
in a
curved spacetime, for instance D-branes in the presence of black hole
then
it is natural to expect that the low energy effective theory will be
given
by the dimensional reduction of the supersymmetric gauge theory coupled
with supergravity in the ten dimensional curved spacetime.

Let us consider the supersymmetric Yang-Mills theory in the
$D$-dimensional
space-time with  metric $g_{MN}$. The action is
\begin{eqnarray}
\label{M1}
I=\int d^Dx\sqrt g tr\{-\frac{1}{4}F_{MN}F_{PQ}g^{MP}g^{NQ}
-\frac{1}{2}{\bar\theta} \Gamma ^MD_M\theta\}
\end{eqnarray}
where $F_{MN}=i[D_M,D_N],~~ D_M=\nabla_M -iA_M$. Let $\gamma :x^M=
x^M(\sigma),~
\sigma=(\sigma_0,...,\sigma_p)$ be a $p+1$-dimensional submanifold
and let us consider the dimensional reduction to $\gamma$.
One has $A_M=(A_{\alpha},~ Y_i), \alpha =0,...,p;~ i=p+1,...,D-1$
and $F_{MN}=(F_{\alpha\beta},F_{\alpha i},F_{ij}),~
g^{MN}=(g^{\alpha\beta},g^{\alpha i},g^{ij})$. The bosonic Lagrangian
is
\begin{eqnarray}
\label{M2}
L=-\frac{1}{2}tr[D_{\alpha}Y_iD_{\beta}Y_jg^{\alpha\beta}g^{ij}
-\frac{1}{2}[Y_i,Y_j][Y_m,Y_n]g^{im}g^{jn}+...]
\end{eqnarray}
Here $Y_i=Y_i(x^P(\sigma)),~g_{MN}=g_{MN}(x^P(\sigma))$.
For a $0$-brane $x^M=x^M(\tau)$ in the gauge $A_0=0,~g^{0i}=0$
one gets the Lagrangian (\ref{1.3}) describing the bosonic part
of M(atrix) theory in curved space.
If one takes $p=1$ then the Lagrangian (\ref{M2})
describes  the matrix string \cite{DVV}
in curved background.

The corresponding Hamiltonian is
\begin{eqnarray}
\label{M4}
H=\frac{1}{2}tr[P^iP^jg_{ij}
-\frac{1}{2}[Y_i,Y_j][Y_m,Y_n]g^{im}g^{jn}]
\end{eqnarray}
One deals with quantum mechanics in the dependent on time background
$g^{ij}(\tau)=g^{ij}(x(\tau))$.
Now the properties of the matrix
quantum mechanics depend on the choice of the curve $x(\tau)$.
The one-loop effective action for the theory
(\ref{M1}) with the  metric $g_{MN}$ can be
computed using the background field method by the standard
procedure. One gets corrections to the phase shift $\delta$
obtained in \cite{BFSS,CT}.
If one takes
geodesics near the singularity then generically one gets the creation of
particles (D$0$-branes)  and there is
back reaction of the gas of D-branes to the metric which
generically is described by the equations:
\begin{eqnarray}
\label{M6}
R_{\mu\nu}-\frac{1}{2}Rg_{\mu\nu}=<T_{\mu\nu}>
\end{eqnarray}
where $T_{\mu\nu}$ is the energy-momentum tensor of D-branes.
One hopes that one can study the singular regime \cite{Ber}
by using the  Matrix theory framework \cite{HP,LM,BFKS}.

Matrix theory Lagrangian (\ref{1.3}) is known to be closely
related to the matrix regularization of supermembrane
\cite{WHN,EMM}. Kappa-symmetry of the supermembrane
in the curved background requires the background be a solution of
eleven dimensional supergravity \cite{BWT}.
We have considered here the metric $g_{\mu \nu}$
in Matrix theory and in matrix string theory as an arbitrary
fenomenological background. This is different from the compactification
prescription discussed in \cite{Sei,Sen}.
Perhaps the metric should be fixed to admit the large N limit.
One expects here a relation with recent considerations
in \cite{Mal,MS,Hya}

$$~$$
{\bf ACKNOWLEDGMENTS}
$$~$$

We are very grateful to Viktor Berezin, Efim Dinin, Valeri Frolov,
Valentin Zagrebnov
and others participants of the Arvilski seminar for extremely friendly
and
stimulating discussions and encouragements.  I.A. and I.V. are supported
in part by  RFFI grants 96-01-00608 and  96-01-00312, respectively.

\end{document}